\documentclass[amssymb,aps,twocolumn,pra,nobibnotes,secnumarabic,prc]{revtex4-1}
\usepackage{graphicx}
\usepackage{lineno}
\usepackage{dcolumn}   
\usepackage{bm}        
\usepackage{amssymb}   
\usepackage{booktabs}
\usepackage{setspace}

\newcommand{\classoption}[1]{\texttt{#1}}

\newcommand{\PbPb}{\ensuremath{\mbox{Pb--Pb}}}
\newcommand{\sqrtSnn}{\ensuremath{\sqrt{s_{\mathrm{NN}}}}}
\newcommand{\pt}{\ensuremath{p_{\mathrm{T}}}}
\newcommand{\fq}{\ensuremath{F_{\mathrm{q}}}}
\newcommand{\fqe}{\ensuremath{F_{q}^{e}}}

\expandafter\ifx\csname package@font\endcsname\relax\else
 \expandafter\expandafter
 \expandafter\usepackage
 \expandafter\expandafter
 \expandafter{\csname package@font\endcsname}%
\fi
\DeclareRobustCommand\substyle{\name@idx{document substyle}}%
\DeclareRobustCommand\classoption{\name@idx{document class option}}%
\DeclareRobustCommand\classname{\name@idx{document class}}%
\def\name@idx#1#2{%
 {\ttfamily#2}%
 \index{#2\space#1=\string\ttt{#2}\space#1}\index{#1>#2=\string\ttt{#2}}%
}%

\begin{document}
\widetext
\title{Scaling Properties of Multiplicity Fluctuations in the AMPT Model}
\author{Rohni Sharma and Ramni Gupta}
\email[ email:]{ramni.gupta16@gmail.com}
\affiliation{Department of Physics, University of Jammu, Jammu \& Kashmir, India}
\begin{abstract}
From the events generated from the MC code of a multi-phase transport (AMPT) model with string melting, the properties of multiplicity fluctuations of charged particles in $\PbPb$ collisions at $\sqrtSnn=\rm{~2.76 \,TeV}$ are studied. Normalized factorial moments, $\fq$, of spatial distributions of the particles have been determined in the framework of intermittency. Those moments are found in some kinematic regions to exhibit scaling behavior at small bin sizes, but not in most regions. However, in relating $\fq$ to $F_{2}$ scaling behavior is found in nearly all regions. The corresponding scaling exponents, $\nu$, determined in the low transverse momentum ($\pt$) region $\le$ 1.0 GeV/c are observed to be independent of the $\pt$ bin position  and  width. The value of $\nu$ is found to be larger than 1.304, which is the value that characterizes the  Ginzburg-Landau type second order phase transition. Thus there is no known signature for phase transition in the AMPT model. This study demonstrates that, for the system under investigation, the method of analysis is effective in extracting features that are relevant to the question of whether the dynamical processes leading phase transition are there or not. 
\end{abstract} 
\maketitle
\section{Introduction}
Critical phenomenon is studied in many areas of physics because of its property of universality. In condensed matter near critical temperature the tension between the collective and the thermal interactions results in clusters of all sizes: thus scale-independent clustering is one of the observable signatures of critical behavior~\cite{binny}. Scaling properties of fluctuations in spatial distributions are hence studied to learn about the dynamics of the systems. In the field of heavy-ion collisions, specific measures for detecting scaling properties were proposed \cite{ref:hn, ref:ising} making use of the Ginzburg-Landau (GL) theory of second-order phase transition (PT) and of the Ising model to simulate spatial patterns. Critical behavior of a system undergoing phase transition has the property that it exhibits fluctuations of all scales. Detection of the scaling properties of local multiplicity fluctuations in the particle production in heavy-ion collisions has been proposed as a signature of the critical phenomenon~\cite{hy,ahep14,ar16}. As observed in cosmic ray event~\cite{ref:cosmic} and in high energy collision experiments~\cite{ref:exp1,ref:exp2} large nonlinear fluctuations exist in the process of space-time evolution of high-energy collisions. Such fluctuations are quantified by the use of normalized factorial moments on the bin multiplicities of particles produced in the phase space of variables of interest, that is subdivided into a large number of bins~\cite{ref:abialas,ref:c,ref:kittlebook}, an analytical tool that we shall review later. 
\par
In high energy collision experiments momentum of the produced particles is usually expressed in terms of ($\eta,\phi,p_{T}$), where  $\eta$ is the pseudorapidity, $\phi$ is the azimuthal angle and $p_{T}$  is the transverse momentum~\cite{ref:kittlebook,raghu}. Local properties in the ($\eta,\phi$) phase space of a single event smear if integrated over all $\pt$, whereas narrow $\pt$ intervals give spatial patterns of emitted particles at approximately different times~\cite{hy,ahep14,ar16}. To extract an important information hidden in the two-dimensional distribution of particles, a study in the ($\eta,\phi$) space in the narrow $\pt$ intervals is performed. For any small $\pt$ interval in ($\eta,\phi$) phase space, to learn about the dynamics of the  system through fluctuation study of the multiplicity distributions,  high multiplicities are required.  Successful model calculations of scaling indices indicative of quark-hadron phase transition also point to the need for high-resolution data that can provide information on the local multiplicities at very small bin sizes. After the first proposal~\cite{ref:abialas} to use normalized factorial moments in small bin sizes and beyond, a large number of efforts were put to understand the dynamical fluctuations in the nuclear collisions~\cite{ref:c,z0,ref:edward1}. Due to insufficient high collision energies the total multiplicities were not high enough to populate the bins and hence to avoid substantial averaging. With the availability of high multiplicities at LHC, it is possible to get detailed studies of the  local properties in ($\eta,\phi$) phase space for narrow $\pt$ bins in a single event and carry out event-by-event analysis over a long-range revealing the scaling behavior. 
\par
Here we study the scaling properties of fluctuations in the momentum-space configurations of particles generated in the a multi-phase transport (AMPT) model (Ampt-v1.25t3-v2.25t3). The AMPT model has been quite useful in understanding some of the experimental results~\cite{ampt:ref1,ampt:ref2,ampt:ref3}. It offers the description of the rapidity distribution, transverse momentum spectrum, elliptic flow and $\pi$ correlations of the systems created at RHIC and LHC energies. We investigate the AMPT model for the fractal behavior which manifests in the form of power-law scaling of multiplicity fluctuations with increasing resolution of phase space, known as {\it intermittency}~\cite{ref:abialas}. We also determine scaling exponents, $\nu$, using the normalized factorial moments, determination of the numerical value of which at LHC energies  is strongly argued~\cite{hy,ahep14,ar16}. Low-$p_{T}$ region, where the partons have more time to interact and equilibrate has been investigated for the study of dependence of $\nu$ on  small $\pt$ bins and on the $\pt$ bin width. Any observation of the scaling or/and signal of phase transition would be of interest. It is known~\cite{ref:c,ref:kittlebook,ref:opal1,ref:opal2} that at low energies and small systems MC studies  were unable to describe hadronic data. Results obtained here can be compared with those from the data and may answer some of the questions of hadron production and hence the dynamical processes leading to the quark-hadron phase transition.
\par
This paper is organized as follows. In Sec.\ II a brief introduction to the AMPT model is given. The methodology of analysis is given in Sec.\ III.  Observations and results of the analysis are discussed in Sec.\ IV followed by a summary of the present work in Sec.\ V.
\section{ A brief Introduction to AMPT}
\begin{figure*}[ht]
\centering
\begin{minipage}[t]{0.48\linewidth}
{\includegraphics[height=2.4in,width=3.0in]{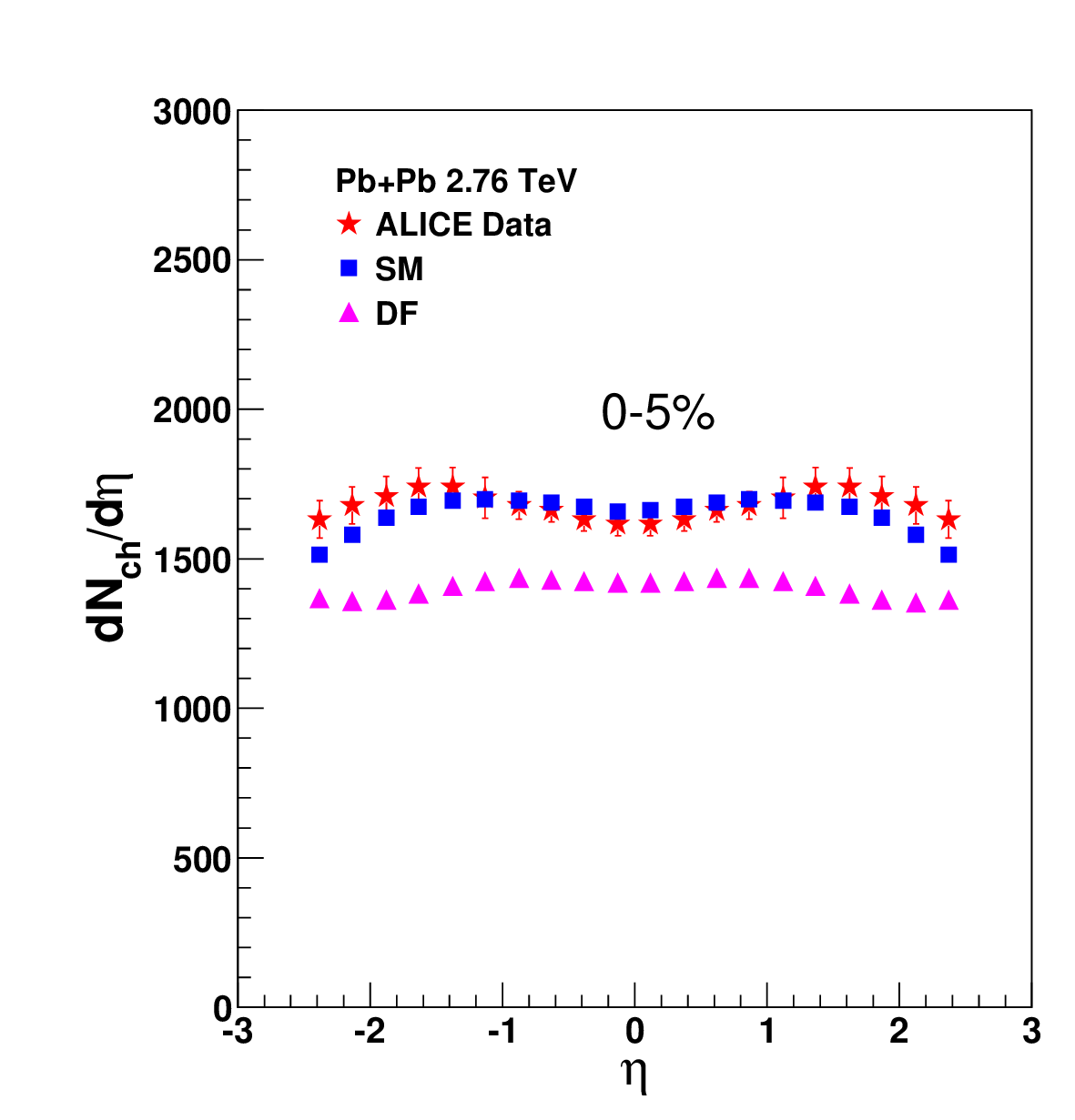}}
\caption{(Colour Online)  Charged particle pseudorapidity density distributions for events generated with DF and SM AMPT events, compared with that from the ALICE data~\cite{alice1}, for the Pb-Pb collisions at $\sqrtSnn = 2.76~\rm{TeV}$ }
\label{figalice1}
\end{minipage}
\begin{minipage}[t]{0.48\linewidth}
{\includegraphics[height=2.2in,width=2.8in]{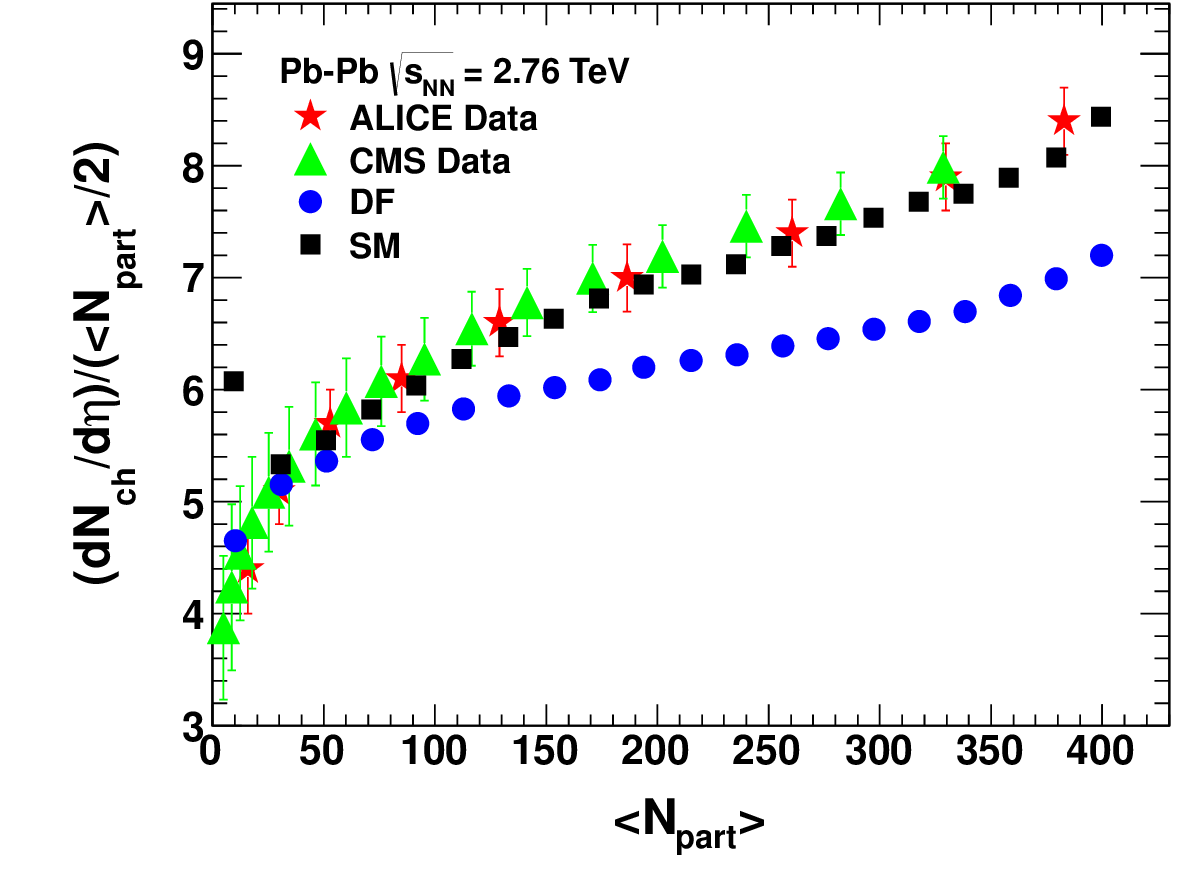}}
\caption{(Colour Online) Dependence of charged particle density on number of participants (centrality) for the DF and SM AMPT events compared with the ALICE~\cite{alice2} and CMS  data~\cite{cms}, for the Pb-Pb collisions at $\sqrtSnn = 2.76~\rm{TeV}$}
\label{figalice2}
\end{minipage}
\end{figure*}
The AMPT model~\cite{ampt:ref5} is framed to study the nuclear collisions lying in the centre of energy range from 5~GeV to 5.5~TeV. It is a hybrid model that includes four main parts: the initial conditions, partonic interactions, hadronization and hadron rescattering. The model exists in two versions: the default (labeled here as DF) AMPT and the string melting (labeled here as SM) AMPT model, depending on how the partons get hadronized. 
\par
The AMPT model that was constructed to simulate relativistic heavy ion collisions, consisting of fluctuating initial conditions from the Heavy Ion Jet INteraction Generator (HIJING) model~\cite{ref:hij}, the elastic parton cascade model viz., the Zhang's Parton Cascade (ZPC)~\cite{ref:zpc}, the Lund string model for hadronization and the A Relativistic Transport (ART)~\cite{ref:art} hadron cascade, is now called the default version. DF AMPT is essentially a string and minijet model with HIJING event generator used for producing initial strings and minijets.  In the DF version hadronization follows the Lund string fragmentation model, in which when the partons stop interacting they recombine with their parent strings and resulting strings are converted into hadrons. Whereas this model was able to reasonably  describe the rapidity distributions and $p_{T}$ spectra in heavy ion collisions from CERN SPS to RHIC energies, it underestimated  the elliptic flow observed at RHIC energies. The string melting version of the model was constructed with all excited hadronic strings in the overlap volume being converted into partons, that was not included in the parton cascade of the DF model~\cite{ref:reason}. Thus the string melting AMPT model consists of fluctuating initial conditions from the HIJING model which converts produced hadrons to quarks and antiquarks, the elastic parton cascade ZPC, a quark coalescence model for hadronization which combines two quarks(antiquarks) into mesons and three quarks(antiquarks) into baryons(antibaryons). As partons freeze-out dynamically at different times in the parton cascade, hadron formation from their coalescence  occurs at different times, leading to the appearance of a co-existing phase of partons and hadrons during hadronization. The final hadron transport  is done via the ART model. 
\par
We generated event samples using the DF and the SM modes of the AMPT model  for the $\PbPb$ collisions at $\sqrtSnn=\rm{2.76~TeV}$ with model parameters similar to that in~\cite{ref:dronika} where a good description of the multiplicity density at all energies from $\sqrtSnn=\rm{7.7~GeV}$  to $\rm{2.76~TeV}$ has been obtained. For Lund string fragmentation parameters $a= \rm{2.2}$, $b=0.5 \, (\rm{GeV^{-2}}$), for the parton scattering cross section screening mass $\mu=1.8$ $\rm{(fm^{-1})}$ and  QCD  coupling constant $\alpha_{\rm{s}} \rm{= 0.47}$ have been used.  The initial and final state radiation in HIJING  has been turned off.  For the set of parameters used in this work, Fig.~\ref{figalice1} shows the pseudorapidity distributions of the generated charged particles in the 0-5\% most central ($\rm{b < 3.5\,fm}$) sample from the two modes of the AMPT model and compared with the ALICE data~\cite{alice1}. In contrast to what is observed in~\cite{ref:dronika}, the SM AMPT sample, at midrapidity $|\eta| \le 1$, is consistent with the ALICE data in comparison to the DF AMPT sample. Fig.~\ref{figalice2}  displays the dependence of the charged particle multiplicity density at per  participant pair on the number of participants. It is seen that the results from the AMPT model with string melting (solid squares) describes reasonably the ALICE~\cite{alice1,alice2} and the CMS data~\cite{cms} whereas, with exception to the ultra-peripheral case, the DF AMPT  version underpredicts the experimental data. Since the SM version agrees better with the data on $dN_{ch}/d\eta$, it is natural for us to choose that version to generate events for  us to study the event-to-event fluctuations of the spatial distributions of the particles produced in the two-dimensional ($\eta,\phi$) space.
\par
The charged particle low $p_{T}$ spectra is not described by the SM AMPT event sample for the same system and same energy. However, based  on the purpose of the analysis, the model parameters can be constrained  to  explain various experimental distributions. The set of parameters as used in~\cite{zlin}, which reasonably describe both charged particle multiplicity and $p_{T}$ spectra may be used for the similar studies in future.
\section{The Method}
\label{sec2}
We investigate the charged particle multiplicity distributions in the two-dimensional phase space constituted together with the basic variables, pseudorapidity ($\eta$) and azimuthal angle ($\phi$) in the small transverse momentum ($\pt$) intervals~\cite{raghu}, in terms of which particle emission in heavy ion collisions can be analyzed. We use normalized factorial moments to quantify fluctuation properties of bin multiplicities as proposed in~\cite{hy}. The ($\eta,\phi$)  space, to be specified below, will be divided into a square lattice with $M_{\eta} \times M_{\phi}$ bins, $M_{\eta}$ and $M_{\phi}$ being the number of bins along $\eta$ and $\phi$ respectively. Only particles produced in a small $\pt$ interval $\Delta\pt$, will be analyzed for each event although several values of $\Delta \pt \le 1.0 $ GeV/c will be considered~\cite{ar16}. Charged particles generated in an event, in the selected $\eta$, $\phi$ and $\pt$ cuts are mapped onto the ($\eta,\phi$) phase space. The normalized factorial moments for each event $e$ are defined as,
\begin{equation}
F_{q}^{e}(M) = \frac{f_{q}^{e}(M)}{[f_{1}^{e}(M)]^{q}}
\label{eq1}
\end{equation}
where,
\begin{equation}
f_{q}^{e}(M)   =  \langle n_{m}(n_{m}-1)......(n_{m}-q+1)\rangle_{e}
\label{eq3}
\end{equation}
in which the order of the moment, $q$, is a positive integer $\ge  2$, $M = M_{\eta} \times M_{\phi}$ is the number of 2D bins, $n_{m} \ge q$ is the bin multiplicity and $\langle~\ldots~\rangle_{e}$ is the average over all bins for the event $e$.
\\
More explicitly Eq.~(\ref{eq3}) can be rewritten as 
\begin{equation}
f_{q}^{e}(M)= \frac{1}{M} \sum_{m=1}^{M} n_{m}(n_{m}-1) \ldots (n_{m}-q+1).
\label{eq31}
\end{equation}
Thus $f_{1}^{e}(M)$ is just the average bin multiplicity $\langle n \rangle _{e}$ of the $e^{th}$ event. 
\\
Upon averaging over all $N$ events, we obtain $F_{q}(M)$ from Eq.\ (1)
\begin{equation}
F_{q}(M) =  \frac{1}{N}\sum_{e=1}^{N} F_{q}^{e}(M).
\label{eq2}
\end{equation}
The virtue of $F_{q}$ is that it filters out statistical fluctuations and $F_q(M)=1$ if the multiplicity distribution is pure Poissonian~\cite{ref:abialas}.\\
If $F_{q}$ has power law dependence on $M$  as
\begin{equation}
F_{q}(M) \propto M^{\varphi_{q}},
\label{inter2}
\end{equation}
the phenomenon is referred to as {\it intermittency} and is a signature of self-similarity of fluctuation patterns of particle multiplicity  that means lack of any particular spatial scale in the system. This scaling we refer here as {\it M-scaling}. $\varphi_{q}$ is called {\it intermittency index}, a positive number \cite{hy,ahep14,ar16,ref:abialas,ref:rchwa} that characterizes the strength of the intermittency signal. A non-vanishing $\varphi_{q}$, is an evidence for the existence of dynamical fluctuations. \\
Even if the scaling behavior in ~Eq.\,(\ref{inter2}) is not strictly obeyed, it is possible that $\fq$ satisfies the power law behavior
\begin{equation}
F_{q} \propto F_{2}^{\beta_{q}}.
\label{fq}
\end{equation}
Hwa and Nazirov~\cite{ref:hn} found   that in the Ginzburg-Landau description of second order phase transition Eq.~(\ref{fq}) is well satisfied and that the scaling exponent $\beta_{q}$ satisfies the equation
\begin{equation}
\beta_{q} = (q-1)^{\nu}, \qquad \nu=1.304 \
\label{beta}
\end{equation}
where $\nu$ is a dimensionless number. Power law scaling of $\fq$ with $F_{2}$ as defined in Eq.~\ref{fq}  is referred here as F-scaling. $\nu$ specifies the property of scaling and characterizes the system under study. Experimental verification of Eq.~(\ref{beta}) has been observed for optical systems at the threshold of lasing~\cite{ref:photon} but yet to be observed and verified in the heavy ion collision experiments.
\begin{figure*}
   \begin{center}
     \includegraphics[height=2.8in,width=6.in]{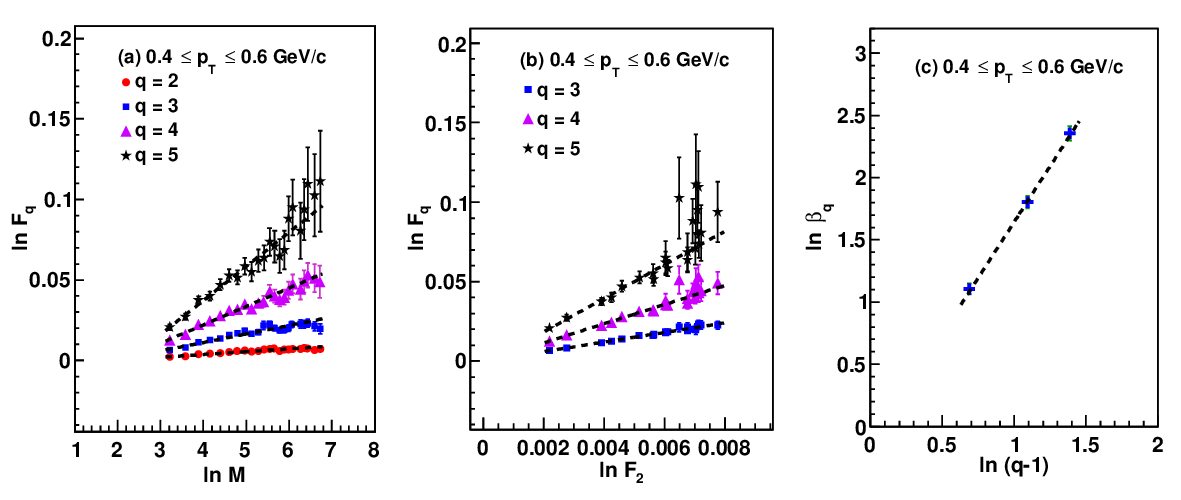}
     \caption{(Colour Online) (a) \it M-scaling behavior of $\fq$. (b) F-scaling behavior of $\fq$ (c)  Log-log plot of $\beta_{q}$ versus ($q-1$), in two dimensional ($\eta,\phi$) phase space in  $0.4 \le \pt \le 0.6$GeV/c bin.}
\label{mscal1}
   \end{center}
 \end{figure*}
\par
Ginzburg-Landau theory is a mean field theory that does not account for spatial fluctuations in a real system. In~\cite{ref:ising}  Cao, Gao, and Hwa investigated the two dimensional Ising model that accounts for spatial fluctuations and determined that $\nu$ depends on temperature (T), one of the control parameters of the model. For critical temperature $T_{c}=2.3 \, \rm{J/k_{B}}$ (Ising parameters), $\nu=1.0$ and that larger $\nu$ occurs at T less than $T_{c}$. In heavy ion collisions where the temperature is not directly measurable quantity, by determining the value of $\nu$ one can infer about the temperature relative to the theoretical critical temperature expressed in terms of parameters in the Ising model. 

\section{Analysis and Observations}  
Local multiplicity fluctuations in the spatial patterns of the events generated using SM AMPT have been examined. A sample of 150K minimum bias  events is generated. A sample of about 10K events in the centrality bin $0 \le b \le 3.5$ fm (corresponding to 0-5\% centrality) has been analyzed. The analysis is performed for charged particles (pions, kaons and, protons) in the phase space region with the kinematic cuts, $|\eta| \le 0.8$ and full azimuthal angle  in the small $\pt$ intervals ($\Delta \pt$)~\cite{ar16}, with $\pt \le 1.0$ GeV/c.  

\begin{figure*}
   \begin{center}
     \begin{tabular}{cc}
     \includegraphics[height=4.0in,width=5.2in]{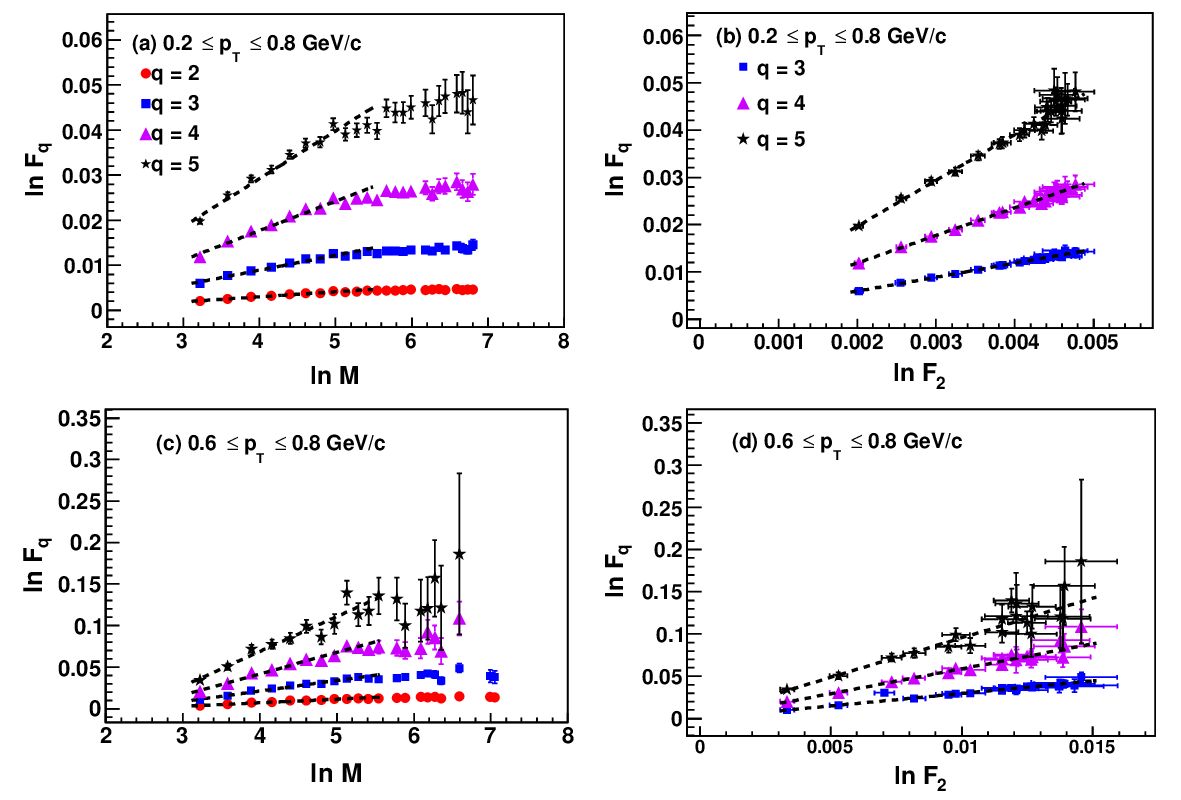}
      \end{tabular}
         \caption{(Colour Online) \label{mscaldiff} \it (a), (c) M-scaling for $q$ = 2,...,5; (b), (d) F-scaling behavior,  in the $0.2 \le \pt \le 0.8$ GeV/c and $0.6 \le \pt \le 0.8$ GeV/c $\pt$ bins respectively }
    \end{center}
 \end{figure*}
\par
For  the study of scaling behavior, we examine the normalized factorial moments $\fq (M)$ (writing simply $\fq$ now onwards) vs $M$ in log-log plots for $q$ =2,...,5. Event factorial moments ($\fqe$) are determined using Eq.\,(\ref{eq1}) for the charged particle density distributions in the ($\eta,\phi$) phase space with partitioning using $M$ = 5 to 30 along the two dimensions.  Fig.~\ref{mscal1} (a) shows the log-log plot of $\fq$ with $M$  in $\pt$ bin $0.4 \le \pt \le 0.6$ GeV/c, for the order parameters q = 2, 3, 4 and 5. We observe an increase of $\fq$ with $M$ for all different $q$'s depicting a self-similar generation of charged particles in this case. A straight line fit to the plots as shown in the figure, gives intermittency index, $\varphi_{q}$. Non-zero values of the intermittency indices are obtained for all $q$, indicating the presence of fluctuations of non-statistical nature in the spatial distribution of charged particles generated by the AMPT model and hence the self-similar fractal structures in this $\pt$ bin.
\par
 A straight line behavior i.e., {\it F-scaling} is observed for $q$ = 3, 4, 5 when $\ln \fq$ is plotted against $\ln F_{2}$ as is shown in Fig.~\ref{mscal1} (b). Error bars on the data points in Fig.~\ref{mscal1} (a) and (b) are the statistical errors, calculated as suggested in~\cite{staterror}. Using Eq.(\ref{fq})  and (\ref{beta}) to describe power $\beta_{q}$, the scaling exponent $\nu=1.79 \pm 0.10$ is obtained by the straight line fit to the $\ln\beta_{q}$ vs $\ln(q-1)$ plot (Fig~\ref{mscal1} (c)).
\begin{figure*}
   \begin{center}
     \begin{tabular}{cc}
          \includegraphics[height=2.3in,width=4.8in]{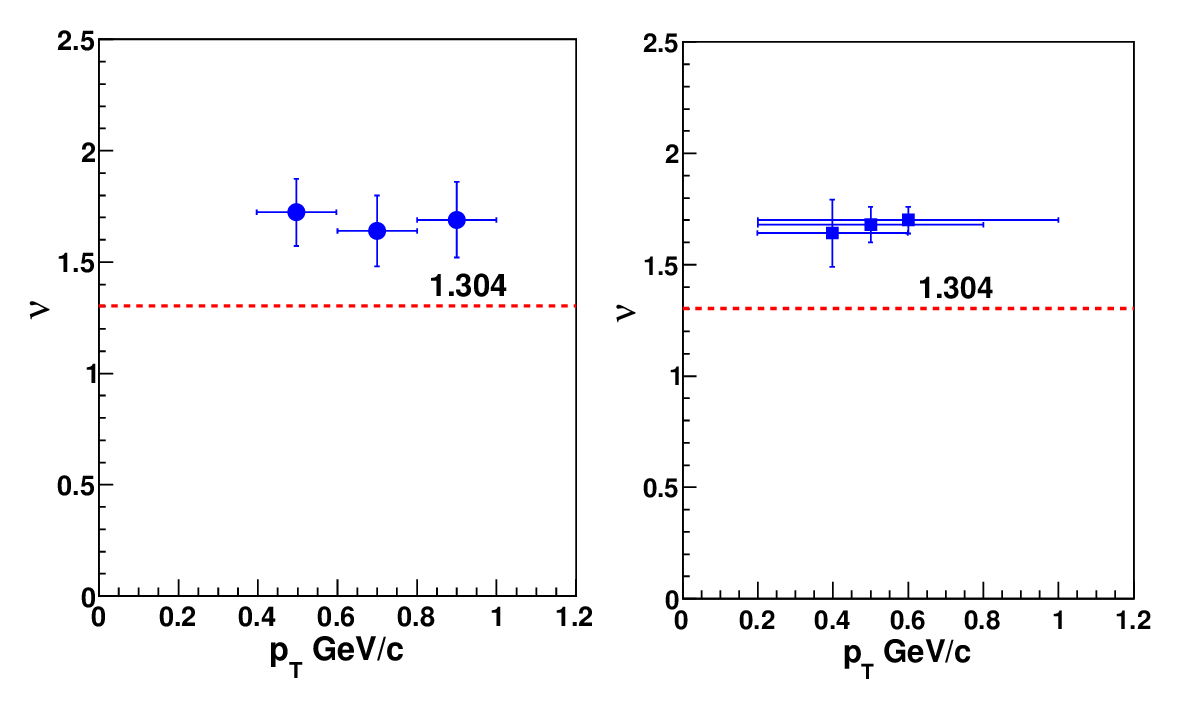}
      \end{tabular}
         \caption{(Colour Online) \label{nuvspt1}\it $\nu$ dependence on $\pt$ bins is shown. The left panel displays the central $\pt$ values, $\pt$ = 0.5, 0.7, 0.9 GeV/c, with the non-overlapping bin width of $\Delta\pt$ = 0.2 shown as horizontal bars. The right panel displays the central $\pt$ values, $\pt$ = 0.4, 0.5, 0.6 GeV/c, per bin with the overlapping bin widths of $\Delta\pt$ = 0.4, 0.6 and 0.8, correspondinly.}
    \end{center}
 \end{figure*}
\par
Similar analysis is performed for $\Delta \pt$ = 0.2 GeV/c at $\pt$ = 0.5, 0.7 and 0.9 GeV/c, and for $\Delta \pt$ = 0.4, 0.6, 0.8 GeV/c at $\pt$ = 0.4, 0.5, 0.6 GeV/c respectively.  Power law scaling of $\fq$ with $M$ cannot be established to be (fully) observed in any of these intervals as can be seen in  Fig.~\ref{mscaldiff} (a) and (c) which show $\ln\fq$ vs $\ln M$ plots for two of these $\pt$ intervals.  $\fq$ observes  power law  for a few low  $M$ values followed by saturation at higher $M$ region and thus no {\it M-scaling}. Now even if {\it M-scaling} is not there to a high degree of accuracy, $\fq$  is observed~\cite{ref:hn} to satisfy {\it F-scaling} (Eq.(\ref{fq})).  For the $\pt$ bins exhibited in Fig.~\ref{mscaldiff} (a) and (c) that show no {\it M-scaling}, the same data points are reproduced in Fig.~\ref{mscaldiff} (b) and (d) that show {\it F-scaling} in the log-log plot of $\fq$ vs $F_{2}$. Scaling exponents ($\nu$) are determined in each of these  $\pt$ bins, as given in Table~\ref{tab2} and are shown in Fig.~\ref{nuvspt1} where $\nu$ values for the average $\pt$  of each  bin are plotted for the above listed non-overlapped and overlapped bins. The dashed line in the plots corresponds to $\nu = 1.304$, the value that is obtained for the second-order phase transition in the GL formalism~\cite{ref:hn}. For the $0.2 \le \pt \le 0.4$ GeV/c  bin no clear conclusions can be drawn about  {\it M-scaling} and  {\it  F-scaling} (not shown). For rest of the $\pt$ intervals, it can be observed that the value of scaling exponents lie in the same range within errors. However, they are well above 1.304. This result, therefore, gives a strong implication that the AMPT model does not simulate events that contain the properties of phase transition as that in the Ginzburg-Landau formalism for second-order phase transition.
\par 
We emphasize that this study demonstrates the effectiveness of the analysis in determining whether spatial multiplicity fluctuations contain any evidence for the phase transition. It is therefore natural at this point to suggest that this method of analysis should be applied to the real data from the experiments at LHC in order to ascertain whether quark-hadron phase transition in the GL mode actually takes place in nature.
\begingroup
\setlength{\tabcolsep}{4pt}
\squeezetable
\begin{table}\centering
\caption{Scaling exponents in the various $\pt$ intervals.}
\label{tab2}
\begin{ruledtabular}
\begin{tabular}{c c} 
{$\pt$} &  {$\nu$ }  \\
\textbf{(GeV/c)} &    \\
\hline \hline
$0.2 \le \pt < 0.4$       &  --- 		     \\    
$0.4 \le \pt < 0.6$       &  1.72 $\pm$ 0.15 	    \\
$0.6 \le \pt < 0.8$       &  1.64 $\pm$ 0.16 	   \\
$0.8 \le \pt < 1.0$       &  1.69 $\pm$ 0.17 	     \\
$0.2 \le \pt < 0.6$       &  1.64 $\pm$ 0.14 	   \\
$0.2 \le \pt < 0.8$       &  1.68 $\pm$ 0.07 	     \\
$0.2 \le \pt \le 1.0$       &  1.70 $\pm$ 0.06 	     \\
\end{tabular}
\end{ruledtabular}
\end{table}  
\endgroup
\section{Summary}
We have studied local multiplicity fluctuations in the spatial patterns of the generated charged particles in the string melting version of the AMPT model  for $\PbPb$ collisions at $\sqrtSnn=2.76$ TeV, using the self-similar analysis of normalized factorial moments for order parameter $q$ = $2$ \rm{to} $5$ in the two dimensional ($\eta,\phi$) phase space in small $\pt$ bins.  Charged particle generation in the model does not show {\it M-scaling} except in the bin $0.4 \le \pt \le 0.6$ GeV/c. However, {\it F-scaling} is  observed to exist in almost all $\pt$ bins. We measured parameter $\nu$, which characterizes the intermittency indices derived in particular analysis. The value of $\nu$ is observed to be independent of the $\pt$ bin and the $\pt$-bin width. The value of the scaling index $\nu$ is different from that of 1.304, a value obtained for the second order phase transition in the Ginzburg-Landau formalism. 
However, for the AMPT model, that describes the single particle LHC data, the observed effects are foreseen to be observed in the experimental data.  Thus results obtained here should be checked by the similar analysis of the experimental data. In case of disagreement of the results between the two, that would require modification in the AMPT which has been tuned to agree with the data but not with the fluctuations in the bin multiplicities. 
\begin{acknowledgements} 
 The  authors are grateful to Rudolph C. Hwa and Edward K. Sarkisyan-Grinbaum for valuable comments and suggestions for the completion of this work. We acknowledge the services provided by the grid computing facility at VECC-Kolkata, India for facilitating to perform a part of the computation used in this work. 
\end{acknowledgements}

\end{document}